\documentclass[preprint,showpacs,preprintnumbers,amsmath,amssymb]{revtex4-1}

\usepackage[dvips]{graphicx}% Include figure files
\usepackage[monochrome,dvips]{color}
\usepackage{dcolumn}% Align table columns on decimal point
\usepackage{bm}% bold math
\usepackage{array}
\usepackage{caption}

\begin{document}

\bibliographystyle{apsrev4-1}

\newcommand{\units}[1]{\ensuremath{\rm ~#1}}
\newcommand{\cmvs}{\units{cm^2/Vs}}
\newcommand{\mum}{\units{\mu m}}
\newcommand{\cent}{\units{^\circ C}}
\newcommand{\vcm}{\units{V/cm}}
\newcommand{\K}{\units{K}}
\newcommand{\ahat}{\units{\hat{a}~}}
\newcommand{\bhat}{\units{\hat{b}~}}
\newcommand{\chat}{\units{\hat{c}~}}
\newcommand{\ang}{\units{\AA}}

%\preprint{APS/123-QED}

\title{A completely cofacial organic semiconductor}

\author{Brett Ellman}
\affiliation{Department of Physics, Kent State University}
\email{bellman@kent.edu}

\author{Robert J. Twieg}
\affiliation{Department of Chemistry, Kent State University}

\date{\today}% It is always \today, today,

\begin{abstract}
Crystals of 1,3,5-tripyrrolebenzene (TPB) contain closely packed, {\it perfectly cofacial}
stacks of benzene rings with large wavefunction overlap, making it an interesting candidate
organic semiconductor.  We study TPB using a variety of {\it ab-initio} and band-structure
techniques, and find very large $\pi$ overlap in the benzene stacks, broad bands (especially for electrons), and relatively small binding energies for
polarons of both signs,
making TPB a promising quasi-one dimensional electron-transport agent.
We then explore the sources of the unusual packing in TPB, finding that calculations of intermolecular interactions using dispersion-corrected density functional
theory provide valuable insights into why the crystals contain {\it perfectly} cofacial $\pi$-networks.
\end{abstract}

\maketitle
\section{Introduction}
The physics of charge transport in organics is complicated by narrow bands, the formation of tightly bound polarons, dynamic
disorder\cite{coropceanu07} and other phenomena largely absent in conventional semiconductors.
A common determinant of  charge 
mobility in both hopping and band models of conduction is the degree to which
the wavefunction of one molecule
overlaps its neighbors, be they in a crystal, liquid crystal, or organic glass.
Numerous attempts have been made to engineer compounds and supramolecular structures to
maximize the transfer integral\cite{cockroft05,williams93}.  One approach exploits
electrostatic interactions to cofacially stack (i.e., with aromatic rings directly facing to each other)
{\it two types} of complimentary molecules A and B, e.g., a hydrocarbon and
the corresponding perfluorinated compound.  Structurally, this has been successful
in crystals\cite{collings02,cho05}. A similar approach was also used in
a discotic liquid crystal system\cite{bushby06, arikainen00}.  However, this scheme has a
fundamental {\it electronic} limitation.  In a hopping picture, and assuming that A has a lower LUMO energy than B, an electron will have
to thermally surmount a barrier when hopping from the LUMO of A to that of B.
Ideally, then, we desire single constituent cofacial organic semiconductors.  We have recently
undertaken a comprehensive search of the Cambridge Crystallographic Database (CCDC)\cite{ccdc}
to identify known organic crystals with structures containing {\it exactly} overlapping benzene rings.
\cite{other_paper}.  Surprisingly, very few (<10) examples were found.  We report here on
an highly unusual compound, 1,3,5-tripyrrolebenzene (TPB), uncovered in the course of this search.
With the
exception of a crystallographic work\cite{thallapally01}, TPB appears to be unknown to the chemistry
and physics communities.

\section{Techniques}
Band structure calculations were carried out with the density functional theory(DFT) Gaussian basis set code
Crystal03\cite{crystal} using the 6-21g* basis set, B3LYP hybrid functional, and
a 10x10x10 Monkhurst integration net.  {\it Ab-initio} calculations on dimers were performed using
the Gaussian 03 and 09 program suites\cite{gaussian}.  Dimer calculations of the transfer integral used
MP2 theory and the 6-311g** basis set.  The intra-stack dimer geometry
was taken directly from the CCDC database without further optimization.
DFT interaction energies calculated using basis set superposition error techniques used the
6-311++g** basis set with the dispersion-corrected
wb97xd functional\cite{chai08}.  The accuracy of these results was checked using a smaller set of MP2 computations.
Reorganization energies for cations were computed using DFT with the B3LYP functional and 6-31g** basis set.  For anions, we used the 6-31++g** basis
set since diffuse basis functions are essential for calculations on negatively charged molecules.

\section{Results}
\begin{figure}[floatfix]
\includegraphics[width=4 in, clip]{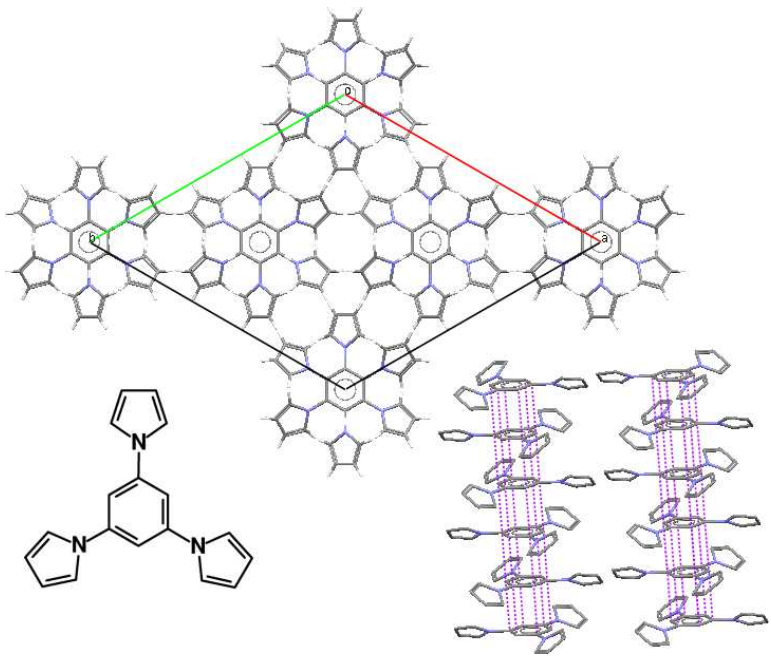}
\caption{\label{struc} Top: The crystal structure of TPB viewed along the hexagonal \chat(top) direction.
The apparent 6-fold symmetry is due to stacks of three-fold molecules alternately rotated by
$60^\circ$.
Bottom: A view showing two stacks.  Note the close contacts (dashed lines) between benzene carbon atoms.
Hydrogen atoms are omitted for clarity.
Inset: 1,3,5-Tripyrrolebenzene structure, omitting hydrogens.}
\end{figure}

TPB crystallizes in the
trigonal space group $R\bar{3}c$ (\#167) with lattice parameters
$a=b=19.42 \ang$, $\gamma=120^\circ$ (hexagonal system) or $a=b=c=11.431 \ang$
and $\alpha=\beta=\gamma=116.305 ^\circ$ (rhombohedral system)\cite{thallapally01}.  From hereon, crystallographic data will refer to the
non-primitive hexagonal system.  We have verified this structure using single-crystal x-ray diffraction on
an independently prepared sample\cite{scott}. We note here that this substance is relatively simple to synthesize, is stable under normal laboratory conditions,
and does not appreciably decompose at its melting point\cite{jarrod_unpub}. The crystal packing,
shown in figure \ref{struc}, has two notable features.  Firstly, the molecules stack in
columns along the \chat direction, with adjacent molecules rotated by $60^\circ$ and
the pyrrole groups rotated out of the benzene ring plane in a ``propeller-like'' fashion.  Molecules in adjacent stacks are offset
along \chat by 1/6 of a lattice vector.
The benzene rings in each stack, however, are {\it exactly} cofacial, i.e., the ring centroids exactly lie over each other and
the ring planes are precisely parallel.
This is in strong contrast to the vast majority of organic crystals, where aromatic rings are 
either offset and/or tilted with respect to one another, often in ``herringbone'' or slip-stacked arrangements\cite{desiraju89}.
Since deviations from perfectly cofacial structures
can have major consequences for transport\cite{coropceanu07}, the exact alignment found in TPB is potentially crucial.

A second feature of the crystal is that the benzene-benzene intermolecular
distance is $3.338 \ang$: this is very short\cite{other_paper} for extended exactly cofacial structures in
crystals (indeed, it is the shortest separation that we found for exactly cofacial, aligned benzene
rings in our search of the CCDC).  On the other hand, the stacks are oriented edge-on to each other and are
relatively widely spaced in the ab-plane (closest C-C distance = 4.224 $\ang$), without
obvious opportunities for $\pi$-wavefunction overlap amongst pyrroles.  This observation
is validated by the band structure calculations, below.

Further discussion is organized as follows.  First, we use dimer quantum chemical and DFT band structure calculations to show that the perfectly stack benzene rings dominate dominate electron transport.  We then compute cation and anion energies to show that polaron binding energies are favorable for transport.  Finally, we use dispersion-corrected DFT calculations to gain insight into the reasons TPB stacks in such an unusual fashion, with perfectly aligned benzene rings.

{\it Ab-initio} MP2 calculations on a molecular dimer cut from
a stack along \chat clearly show the dominance of benzene ring $\pi$-orbitals in the LUMO, and

\begin{figure}[floatfix]
\includegraphics[width=3 in, clip]{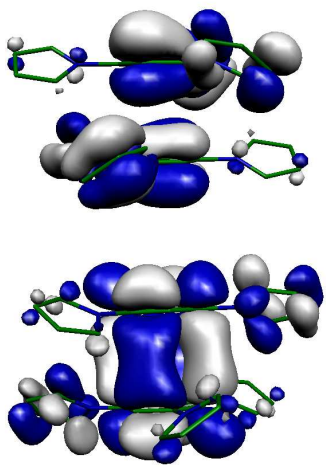}
\caption{\label{dimer}  Top: A HOMO isosurface for a TPB dimer (density=$\units{0.02 ~ electron/Bohr^3}$.  The color corresponds to the phase of the wavefunction.
Note the $\sigma$ orbital-like character of the wavefunction on the benzene ring and the
sizable densities near the pyrrole rings.
Bottom:  The electron density of the a LUMO orbital for TPB.  Note the
$p_z$-like character (with $\units{\hat{z}}$ orthogonal to the benzene ring), large overlap of the wavefunction between the benzene rings, and the relatively
low density about the pyrroles.  Also note the bonding (as opposed to anti-bonding) nature of the interactions.
}
\end{figure}

of benzene $\sigma$ and various pyrrole orbitals in the HOMO wavefunctions (Fig. \ref{dimer}).
The qualitative importance of the benzene
ring for the empty LUMO levels, and of the pyrrole rings for the HOMO, are clear.
Since the valence and conduction bands of an organic semiconductor are formed, in a tight-binding sense,
from the HOMO and LUMO, respectively, we expect that electron conduction will be dominated by a few
wavefunctions from the exactly cofacial, closely-packed benzene rings, leading to large electron bandwidths.
On the other hand, we expect many, individual, relatively narrow valence bands supporting hole transport.

\begin{figure}[floatfix]
\includegraphics[width=4 in, clip]{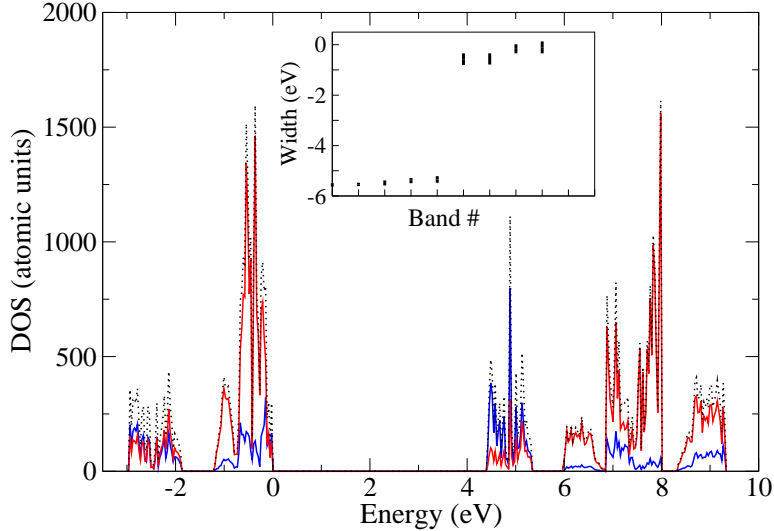}
\caption{\label{dos} Density-of-states computed for a TPB crystal.  The DOS are
in atomic units ($\units{electron/(Bohr^3~Hartree)}$).  The zero of energy corresponds to the
valence band edge.  The dotted line is the total
DOS, the blue line is the contribution from atoms in the central benzene group, and the red curve
is the DOS from the pyrrole rings.  Note, in particular, the relatively large contribution
from the cofacial benzenes to the DOS near the conduction band edge.  Inset: Bandwidths for TPB
for states near the Fermi level.  The conduction band is significantly broader than the individual bands
near the Fermi energy, due to the close benzene-benzene packing and resulting  wavefunction overlap. }
\end{figure}

\begin{figure}[floatfix]
\includegraphics[width=4 in, clip]{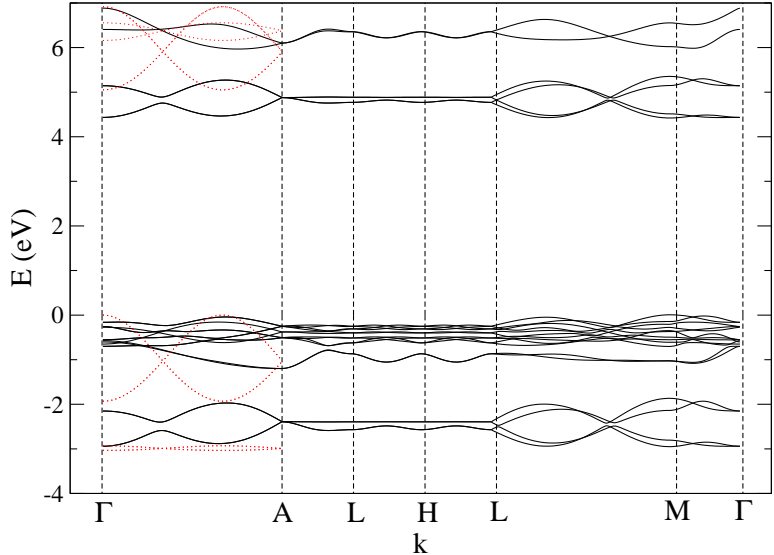}
\caption{\label{band}  Band structure near the Fermi energy in TPB.  The k-space labels $\Gamma$, A, L, H, and M follow the {\it hexagonal}
convention and correspond to (000), (001), (101), (111), and (010), respectively.  The zero of energy corresponds to the valence band top.  
Note the broad bands along the stacking axis
(001), and the much narrower, flatter bands in the orthogonal directions.  The zero of energy is the Fermi energy.
The dotted curves from $\Gamma$ to A are the equivalent bands
for a "TPB benzene crystal", in which the pyrroles are replaced with hydrogen - see text for discussion.
}
\end{figure}

These conclusions are born out by band structure calculations.  Figure \ref{dos} shows the
total density
of states for TPB along with the components of the DOS ascribable either to the atoms of the benzene
ring (including the hydrogens) or to the pyrrole groups.  The valence states (which support hole transport) are complicated, with individual bands being
quite narrow (though the overall bandwidth due to overlap is fairly broad - see below).
The DOS is clearly dominated
by the states originating from the pyrrole rings, consistent with the dimer calculations.
The conduction band states are more
interesting.  Considering the \chat data in Fig. \ref{band}, the individual lowest energy bands are seen
to be quite broad (about 460 meV).  Furthermore, the partial DOS in Fig. \ref{dos} shows that
the electron transport states {\it nearest} the band edge are almost entirely due to the benzene rings.
As emphasized in, e.g., Ref. \cite{cheng03}, it is precisely these states (within several $k_B T$ of the band edge)
that are most important for transport.

The conduction band structure is the macroscopic consequence of the $\pi$ overlap in Fig. \ref{dimer}, and convincingly demonstrates
the significance of the exact cofacial structure for transport in TPB.  To quantify this, note that,
in the inset to Fig. \ref{dos}, four bands overlap
both at the top of the valence and the bottom of the conduction bands.  The total bandwidth of the upper valence bands
is 393 meV, comparable to the top two overlapping bands in naphthalene from a similar calculation (337 meV).  However, the conduction bandwidth
is 933 meV, much larger than naphthalene's 347 meV conduction-band value: compared to naphthalene, the benzene
rings in TPB support a large transfer integral.

The bandgap at the $\Gamma$-point is about 4.5 eV,
with slightly smaller values ($\sim$ 4.4 eV) elsewhere in the zone (the standard issues\cite{yakovkin07, morisanchez08} concerning DFT bandgaps apply).
We have experimentally measured the the
ultraviolet spectrum in the vapor phase of TPB, and find the onset of absorption occurs at 300 nm.  This provides
an experimental optical bandgap of 4.2 eV, in reasonable agreement with the calculation.  As a further check of this important parameter, we have performed a time-dependent DFT (TDDFT) calculation on TPB.  TDDFT is known to be an excellent technique for computing HOMO-LUMO gaps in organics\cite{zhang2007}.  The first excitation energy was found to be 4.3 eV, again in good agreement.
The conduction band effective mass, 
\begin{equation}
\frac{1}{m^*}=\frac{1}{\hbar^2}\left ( \frac{\partial^2 E}{\partial k^2} \right ) ^{-1},
\end{equation}
computed via numerical differentiation of the DFT bandstructure, is 1.84 $m_e$ at the $\Gamma$-point.  This is relatively low: the corresponding value for
naphthalene along the b-axis, computed in the same fashion, is about 3.5 $m_e$ - again, a favorable factor for
electron transport.
Also, as expected from the highly anisotropic crystal structure, TPB is a quasi-one dimensional electron-transport material:
the conduction band for
k-vectors in the ab-plane (e.g., see Fig. \ref{band}, Brillouin zone points M to $\Gamma$) is much narrower than along $\chat$.

When studying the electronic properties of TPB, it is tempting to ignore the pyrrole groups entirely and consider TPB to electronically be analogous
to close-packed
exactly cofacial stacks of benzene rings widely spaced from each other.  To explore this model, we computed the band structure of a TPB ``crystal'' with the pyrrole groups
replaced by hydrogen, i.e., of a TPB crystal with TPB molecules replaced by benzene molecules.  As expected, the ab-plane bands
have essentially zero width.  However, the c-axis bands, shown in Fig. \ref{band}, are substantially broader than those of TPB and
have a slightly larger bandgap.  One reason for the band narrowing is the delocalization of the LUMO wavefunctions over the pyrrole rings in TPB, resulting in less weight on the benzene ring and a decreased transfer integral.  We might expect this based on the dimer calculations (above) - the LUMO wavefunction, dominantly present near the benzene rings, does have a non-zero presence on the pyrroles.  It is therefore evident that the pyrrole groups play a significant role in both electron and hole transport.  They also are extremely important {\it structurally}, as we will see.

Based solely on the band structure, TPB therefore is a favorable candidate for quasi-one dimensional, high mobility electron
transport.  This conclusion is premature, however, since the mobility may be exponentially suppressed by the polaron binding energy, $E_b$.  In the case of small
polarons, $E_b$ may be approximated by the reorganization energy, $E_b=\lambda_{reorg}/2$.  Following\cite{coropceanu07}, Coropceanu et al.,
$\lambda_{reorg}=\lambda_1 + \lambda_2$ where
\begin{eqnarray}
\lambda_1=&E^{(1)}(M)-E^{(0)}(M)\\
\lambda_2=&E^{(1)}(M^\pm)-E^{(0)}(M^\pm).
\end{eqnarray}
Here, $E^{(0)}(M)$ and $E^{(0)}(M^\pm)$ are the ground-state energies of the geometrically-optimized neutral TPB molecule and the
its anion or cation, respectively.  $E^{(1)}(M)$ and $E^{(1)}(M^\pm)$ are similar, except that the energy of the neutral molecule
is computed using the ion geometry and vice-versa.  On carrying out these calculations, we find for the cation (pertinent to hole transport), $E^+_b$=65 meV.  This may be
compared to naphthalene, for which we compute a
value of 93 meV using the same model chemistry and basis set (in close agreement with the calculation of ref. \cite{coropceanu07}.  Therefore, dressed holes may be
expected to be {\it less} strongly bound in TPB than in naphthalene, making it a competent hole semiconductor.
Of greater interest is the tendency of electrons to form polarons in the very broad conduction band of TPB.
Three-fold symmetric systems
like TPB (e.g., systems with degenerate frontier orbitals) are prone to
Jahn-Teller distortions when negatively ionized \cite{coropceanu07}, resulting in large electron/phonon
couplings and therefore large polaron binding energies.  Our calculations give an anionic $E^-_b$=144 meV.  While higher than the
hole binding energy, this still compares favorably with conventional materials like the acenes\cite{coropceanu07}.  Therefore, there appears to be every
reason to expect that TPB is a promising electron transport agent.

\begin{figure}[floatfix]
\includegraphics[width=4 in, clip]{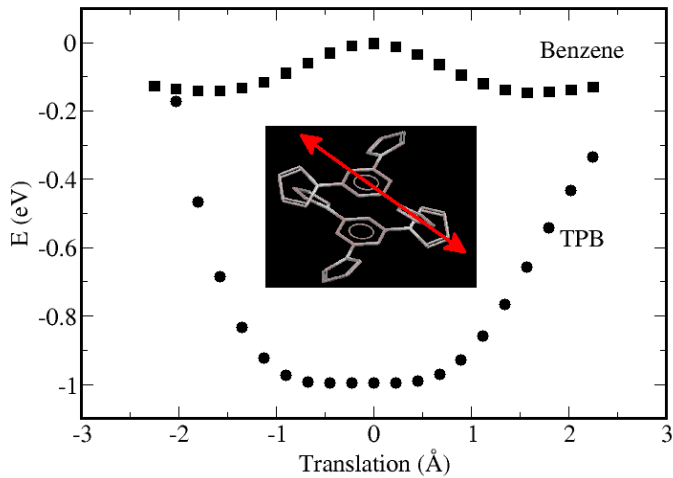}
\caption{\label{trans}  Dimer energy as a function of the relative lateral translation between the molecules for benzene and TPB.  The origin corresponds to the benzene rings lying exactly over each other.  Note the maximum at the origin for Benzene and the broad minimum for TPB.}
\end{figure}

We now address the second question: why does TPB stack in this unusual, exactly cofacial, close-packed fashion?
As noted above, very few compounds indeed crystallize with benzene rings exactly over each other. It is, of course, very difficult to
{\it a priori} predict crystal structures from molecular structures.  We have, however, gained insights from {\it a postori} calculations
of intermolecular interaction energies using dispersion-corrected DFT.
In particular. an incisive clue to the source of the exactly cofacial packing found in TPB comes from studies of the energetic cost of translating a
molecule parallel to the molecular plane within a stack (see the inset to figure \ref{trans}).
(Here we ignore neighboring stacks to isolate the intra-stack contribution.)  Figure \ref{trans}
contrasts the behavior of an intra-stack TPB dimer with a benzene dimer corresponding to the TPB molecules with the pyrroles replaced by hydrogens.
Under the translation shown in the figure, the energy of the benzene
dimer has the expected and long recognized behavior, with a maximum when the rings are over
each other: it is energetically unfavorable for benzene molecules to align exactly cofacially.  The interesting point from the same calculation on TPB molecules is that the energy maximum at the origin
is {\it eliminated}.  The pyrrole rings, either by electronically modifying the central benzene group and/or {\it via} pyrrole-pyrrole
interactions, make it energetically favorable for the molecules to lie directly over each other.  Some of this favorable interaction energy is presumably
electrostatic in nature: Mulliken charge analysis indicates that the pyrrole rings are electron-withdrawing.
Since the central benzene rings in a TPB crystal stack are rotated 60
degrees relative to each other, this results in a negative electrostatic potential energy between adjacent benzene groups (relatively positive carbon atoms lie over relatively negative ones).

One might therefore expect similar packing for other three-fold symmetric molecules
built of a benzene ring with a set of 1,3,5-disposed electron-withdrawing or electron-rich substituents (either of which would break the
six-fold symmetry of the benzene ring).  The reality
here is more complex.  None of the simple 1,3,5-tri-halogenated analogs are as favorably packed as TPB.  However, a number of more complicated
$C_3$-symmetry molecules with benzene cores and large, complex side groups , e.g.,
N,N',N''-tris(1,3-bis(methoxycarbonyl)propyl)benzene-1,3,5-tricarboxamide (CCDC code XEFYEM),
1,3,5-N,N',N''-tris(2-methoxyethyl)trimesic amide (BOHTIB), and 1,3,5-tris(difluoroboronyloxy)-2,4,6-tris((4-fluorophenylimino)methyl) benzene
(XENMEJ, with a boron-containing triphenylene-like core) {\it do} stack exactly cofacially.
Indeed, as discussed in ref. \cite{other_paper}, a large fraction of all
materials with cofacial benzene rings in the CCDC are three-fold symmetric.  We therefore anticipate that an in-depth study of the intermolecular forces between three-fold symmetric compounds with a central phenyl moeity will lead to insights into the enngineering of highly cofacial $\pi$-networks.

\section{Conclusion}
We have shown that the electronic properties of TPB crystals are profoundly influenced by close, perfectly coplanar
columns of benzene rings, particularly for the case of electron transport.  The conduction bandwidth is very large
(compared to our model acene naphthalene) and the effective mass small.
  Polaron binding energies of both signs of charge are also favorable for
transport.  Therefore TPB may well be an excellent candidate for a high mobility, quasi-1d electron semiconductor.  Calculations of the energetics cost of molecular displacements indicate that the pyrrole groups are essential to stabilizing the unusual packing.  These calculations need to be extended to other perfectly cofacial three-fold systems (since the other known systems are significantly larger, this will be computationally expensive).
The electronic properties of other 3-fold
compounds that exhibit cofacial stacking also need to be investigated both theoretically.
Most importantly, experimental measurements of mobility would be fascinating.

\bibliography{paper_sub}

\end{document}